# *MatBase* Metadata Catalog Management

**Christian Mancas***

*Mathematics and Computer Science Department, Ovidius University, Constanta, Romania*

**\*Corresponding Author:** Christian Mancas, Mathematics and Computer Science Department, Ovidius University, Constanta, Romania.

### Abstract

*MatBase* is a prototype intelligent data and knowledge base management system based on the Relational, Entity-Relationship, and (Elementary) Mathematical Data Models. The latter distinguishes itself especially by its rich panoply of constraint types: 61, partitioned into three categories (set, containing 16 types, mapping, containing 44 types, and object) and eight subcategories (general set, dyadic relation, general mapping, autofunction, general function product, homogeneous binary function product, function diagram, and object). They provide database and software application designers with the tools necessary for capturing and enforcing all business rules from any sub-universe of discourse, thus guaranteeing database instances plausibility, a sine qua non condition of data quality. This mathematical data model also includes Datalog, thus making *MatBase* also a deductive, so a knowledge base system. Currently, there are two *MatBase* versions (one developed in MS Access and the other in MS.NET, using C# and SQL Server), used both by two software developing companies and during labs of our M.Sc. students within the Advanced Databases lectures and labs, both at the Ovidius University and at the Department of Engineering in Foreign Languages, Computer Science Taught in English Stream of the Bucharest Polytechnic University. This paper presents *MatBase*'s metadata catalog and its management.

## Introduction

*MatBase* [1-5] is a prototype intelligent Knowledge and Database Management System (KDBMS) built on top of an existing relational DBMS (RDBMS) and based on both the Relational Data Model (RDM) [2,6,7], the Entity-Relationship one (E-RDM) [2,8,9] and the (Elementary) Mathematical one ((E)MDM) [1,2,4,5,10-14], which also embeds Datalog [1,4,6]: its users may define, update, and delete database (db) schemas in any of these three formalisms, which MatBase is automatically translating into the other two ones.

For keeping track of the managed databases, MatBase has its own metadata catalog, made of 373 tables and views, which is based on and managed through the (E)MDM and E-RDM.

## A brief introduction to (E)MDM

(E)MDM schemes are quadruples made from:

- A finite nonempty collection of sets S, partially ordered by inclusion,
- A finite nonempty set of mappings M defined on and taking values from sets of S,
- A finite nonempty set of constraints C over the elements of S and M, and
- A finite set of Datalog programs P associated to the elements of S and M.

S is partitioned into the following four blocks: object, value, system, and computed sets:

- Object ones are partitioned into entity (i.e. atomic) and relationship (i.e. non-functional math relations immune to their domain permutations). In relational dbs (rdbs) they are implemented as tables.
- Value ones are subsets of (programming) data types.
- System sets include at least the data types, the empty set, a distinguished countable set NULLS of null values, and all the sets of the MatBase metadata catalog.
- Computed sets are obtained from all other types of sets by using semi-naïve sets, functions, and relations algebra operators. In rdbs they are implemented as views.

All mappings in M are defined either on object or on computed sets not based on value ones. In rdbs they are implemented as table/view columns. M is partitioned into the following four blocks: attributes, structural functions (implemented in rdbs as foreign keys), system and computed mappings:

- Attributes are taking values from value sets.
- Structural functions from object ones.
- System mappings include unity mappings, canonical projections and injections.
- Computed mappings are obtained from all other types of mappings by using semi-naïve sets, functions, and relations algebra operators (e.g. composition, (Cartesian) product, etc.).

C is partitioned into the following three blocks: set, mapping, and object constraints. Some constraints are associated to corresponding db E-R diagram (E-RD) cycles [8,15]. The simplest cycles are those of length 1: the autofunctions (i.e. of the type f: A → A). Cycles of length greater than one are made from nodes of three types (source, i.e. they are domains of the two mappings that connect them to the cycle, destination, i.e. they are co-domains of the two mappings that connect them to the cycle, and intermediate, i.e. they are the domain for one mapping and the co-domain for the other one that connect them to the cycle) and may only be of the following three types [4,11]:

- Commutative (i.e. having only one source and one destination nodes).
- Circular (i.e. having only intermediate nodes).
- General (i.e. any other cycle than those of types commutative and circular; they have length greater than 3 and at least two sources and two destinations).

The set constraints include 16 constraint types partitioned into two blocks: general set and dyadic relation (i.e. binary math relations defined over a set).

The general set constraints are sub-partitioned into the following five blocks: inclusion, equality, disjointness, union, and direct sum.

Dyadic relation ones (e.g. D ⊂ $S^2$) are sub-partitioned into the following eleven blocks: reflexivity, irreflexivity, symmetry, asymmetry, transitivity, intransitivity, Euclideanity, in Euclideanity, equivalence, acyclicity, and connectivity.

There are 44 mapping constraint types that are sub-partitioned into the following five blocks: general mapping, autofunction, homogeneous binary function product (i.e. of the type f • g: A → $B^2$), general function product, and function diagram.

General mapping constraints are sub-parti¬tioned into the following six blocks: totality (i.e. not null), nonprimeness (i.e. it cannot be part of any key), one-to-oneness (single key), ontoness, bijectivity, and default value.

Autofunction constraints are sub-partitioned into the follow¬ing eleven blocks (autofunctions being particular cases of dyadic relations, for which Euclideanity and connectivity do not make sense, as they would violate function definition): reflexivity, irreflexivity, null-reflexivity (i.e. reflexivity for all not null values), symmetry, asymmetry, null-symmetry (i.e. symmetry for all not null values), idempotency, anti-idempotency, null-idempotency (i.e. idempotency for all not null values), acyclicity, and ca¬no¬ni¬cal subjectivity.

Homogeneous binary function product ones are sub-partitioned into the following fourteen blocks (note that reflexivity does not make sense in this context: why would anybody wish to have two columns of a fundamental table that should store exactly same values?): irreflexivity, null-reflexivity, symmetry, asym-metry, null-symmetry, transitivity, intransitivity, null-transitivity, Euclideanity, in Euclideanity, null-Euclideanity, equivalence, acyclicity, and connectivity.

Function product ones are sub-partitioned into the following three blocks: minimal one-to-oneness (concatenated key), exis¬tence, and nonexistence.

Function diagram constraints include the following ten blocks: commutativity (equality), anti-commutativity (inequality), local commutativity (i.e. corresponding compound autofunction reflexivity), local anti-commutativity (i.e. corresponding compound autofunction irreflexivity), local acyclicity (i.e. corresponding compound autofunction acyclicity), local symmetry (i.e. corresponding compound autofunction symmetry), local asymmetry (i.e. corresponding compound autofunction asymmetry), local idempotency (i.e. corresponding compound autofunction idempotency), local anti-idempotency (i.e. corresponding compound autofunction anti-idempotency), and generalized commutativity (particular case of an object constraint only involving mappings of a same function diagram of type general).

Object constraints are closed Horn clauses (i.e. disjunctions of literals with at most one positive, i.e. unnegated one).

Some of the above constraints are fundamental, while others are derived: for example, set equality is a derived one (from inclusion), just as direct sum (from disjointness and union), equivalence (from reflexivity, symmetry, and transitivity or reflexivity and Euclideanity), totality (from existence), bijectivity (from one-to-oneness and ontoness), etc. are. Dyadic relation ones can always be considered as homoge¬neous binary product ones, where the products are made from their roles (i.e. canonical (Cartesian) projections).

In total, there are only 22 fundamental constraint types in (E) MDM, the remaining 39 being derived. In fact, of course, as all constraints are closed FOPC formulas, only the object constraint is actually fundamental. However, all relevant well-established fundamental math and RDM concepts are considered fundamental in the (E)MDM too (i.e. inclusion, dyadic relation reflexivity, one-to-oneness, minimal one-to-oneness, nonprimeness, ontoness, existence, non-existence, function diagram (conventional) commuta-tivity, anti-commutativity, and generalized commutativity, as well as homogeneous binary function product irreflexivity, symmetry, asymmetry, transitivity, intransitivity, Euclideanity, in Euclideanity, acyclicity, and connectivity).

All five RDBMS provided constraint types are included in (E) MDM too: domain (in co-domain definitions) and referential integrity (from the Key Propagation Principle [2,15,16]) implicitly, while not null (totality), keys (minimal one-to-oneness), and tuple/check (extended to object constraints) explicitly.

To conclude about constraints, always discovering and enforcing all existing ones in the sub-universes modeled by dbs is crucial: any existing constraint that is not enforced in a db scheme allows for storing implausible data in its instances, thus compromising data quality.

### Related work

Lot of work has been published on *MatBase*, the three data models it provides, and Datalog; e.g. see references of this paper, as well as their corresponding references.

### Paper outline

The following two sections introduce the *MatBase*'s metadata catalog structure and management, respectively. The paper ends with conclusion and references.

### *MatBase* metadata catalog structure

As for any DBMS/KBMS, *MatBase*'s metadata catalog is made of a set of fundamental and temporary hidden tables, accessible, through a system password, only to *MatBase* architects and developers, as well as of a set of views (computed sets), accessible to all of its users. Just like for any db it manages, they are partitioned into set categories. For the metadata catalog, these categories are the following: fundamental sets, computed sets, system sets (e.g. the empty set, the NULLS set, etc.), fundamental mappings, computed mappings, system mappings (e.g. cardinal, unity, etc.), constraints, Datalog programs, E-R diagrams (E-RDs), host objects (e.g. used MS Access/SQL Server metadata objects).

The fundamental sets category includes the following four tables: SETS_CATEGORIES (storing the corresponding db, category name, description, etc.), SETS (storing the corresponding category, set names, types, categories, descriptions, cardinal, etc.), REL_SORTS (storing the structure of relationship-type sets, i.e. their canonical Cartesian projections), and RELATIONSHIPS (storing constraints for the particular case of the dyadic relationship-type sets).

The computed sets category includes the following two tables: *SETS (storing data on computed sets, e.g. their math formula, their corresponding SQL one, etc.) and CARTES_PROD_COMP (storing the structure of computed sets of type Cartesian products). Moreover, it includes dozens of views (e.g. for entity-type, relationship-type, value-type, computed-type, system-type sets, for subsets, equal sets, disjoint sets, etc.).

The system sets category includes very many tables: DATABASES (storing paths, names, types, etc. of managed dbs), LANGUAGES (storing the languages for which *MatBase* is localized - currently English, the default one, French, and Romanian only), DICTIONARY (storing equivalent words/phrases in all known languages), OPERATORS (storing all 64 *MatBase* provided operators - be them set, mapping, algebraic, logic, relational, standard math, or text ones), MATH_STRINGS (storing math formulas of computed objects), SQL_STRINGS (storing corresponding SQL ones), FORMS (storing the set of MS Windows forms built upon tables and views, both as part of *MatBase* graphic user interface and as containers of classes mainly including event-driven methods for enforcing non-relational constraints and dramatically enhancing ergonomy for users), NULLS, EMPTY, the system data types (BOOLE, NAT, INT, RAT, etc), etc. It also includes dozens of views (e.g. numeric operators, logic operators, set operators, etc.).

The fundamental mappings category includes only one table, namely FUNCTIONS (storing mapping names, domains, codomains, description, constraints -e.g. totally defined, one-to-one, onto, acyclic, etc.-, default, minimum and maximum plausible values, etc.) and several views (for attributes, structural functions, computed ones, etc.).

The computed mappings category includes tables *FUNCTIONS (storing the subset of computed mappings, their math and SQL formulas, etc.), COMP_FUNCT_COMP (storing the structure of compound mappings), and FUNCT_PRODUCTS (storing the structure of Cartesian function products). It also includes views (e.g. for functions involved in at least one equality, existence, non-existence, etc. constraint).

The system mappings category only includes views (e.g. set-related, mapping-related, logic-related, etc. provided system functions, etc.).

The constraints category contains the largest number of both tables and views. The main table is CONSTRAINTSET (as CONSTRAINTS is a reserved table name in MS Access), which stores corresponding db, constraint names, types, description, associated set, mapping, E-RD cycle (if any), etc. CONSTRAINT_TYPES stores the name, abbreviation, subcategory, implying type (if any), etc. of all 64 constraint types of (E)MDM, plus the two relational ones (domain and referential integrity) that do not need to be explicitly asserted in (E)MDM. CONSTRAINT_CATEGS stores the names of the four constraint categories: set, mapping, object, and relational. CONSTRAINT_SUBCATEGS stores the nine corresponding constraint subcategories: general set and dyadic relation ones (of the set category), general mapping, autofunction, general func-tion product, homogeneous binary function product and function diagram ones (of the mapping category), object, and relational (the only ones in their corresponding categories). IMPLICATIONS stores the pairs of constraints <c, c'>, whenever c implies c'.

Then, there are the tables storing data particular to each constraint type; for example, INCLUSIONS, SET_EQUALS, DISJOINTNESSES, DIRECT_SUMS, UNIONS (storing both sets involved in the corresponding constraint type), EXIST_CNSTR and NON_EXIST_CNSTR (storing (non-)existence constraints data), FUNCT_EQUALS (storing both functions involved in such equalities), etc.

For enforcing constraint sets coherence and minimality [4,17], there are the following seven tables: THEOREMS (storing the type of theorems, which can be either incoherence or redundancy, their name, description, the order in which MatBase applies them, etc.), SCCOHERENCES (storing the non-trivial combinations of set constraints), SCREDUNDANCIES (storing the redundant constraints in such combinations), MCCOHERENCES (storing the non-trivial combinations of mapping constraints), MCREDUNDANCIES (storing the redundant constraints in such combinations), HBRCCOHERENCES (storing the non-trivial combinations of homogeneous binary function products and dyadic relation constraints), and HBRCCREDUNDANCIES (storing the redundant constraints in such combinations).

Besides other constraint-related fundamental tables, this category also includes dozens of views (e.g. *BijectivityConstraints, *CanonicalInjections, *CanonicalSurjections, etc.).

The Datalog category includes lot of fundamental tables and views as well. Main tables in this category are PREDICATES (storing predicate name and types -either intensional or extensional, etc.), INF_RULES (storing the inference rules of Datalog programs), INF_RULES_COMP (storing the structure of the inference rules' bodies), PROGRAMS (storing the name, description, type -either system or user-, and associated relational algebra (RA) equation systems of Datalog programs), PROG_COMPS (storing the inference rules out of which Datalog programs are made of), EXPRESSIONS (the set of algebraic expressions that are part of inference rules), CONSTANTS, VARIABLES, TERMS, ATOMS, FORMULAS (storing the corresponding components of algebraic expressions, respectively), RA_EXPRS (storing the subset of RA expressions), RAEQUATIONS (storing the RA equations corresponding to Datalog inference rules), RAEQSYSTEMS (storing the RA equation systems corresponding to Datalog programs), and RAESCOMP (storing the structure of RAEQSYSTEMS, i.e. the RA equations that are making them up).

Among the views from this category, the most used ones are those of the subsets of extensional and intensional predicates, respectively.

The E-RDs category includes the fundamental tables DIAGRAMS (storing the corresponding dbs, names, descriptions, corresponding jpg file names and paths of the E-RDs, etc.), DIAGRAM_O_COMP (storing the object sets that are part of E-RDs, their position, sizes, etc.), DIAGRAM_A_COMP (storing the structural functions that are arrows of E-RDs, their position, sizes, etc.), and ELLIPSES (storing the attributes that are part of E-RDs, their position, sizes, etc.). Then, there are temporary tables filled when E-RD cycles are detected [18], e.g. ERDCycles, ERDNodes, ERDEdges, etc. There are also views in this category, e.g. ERDCyclesComposedMappings, ERDEdgesDuplicates (showing all E-RD cycles in which edges are part of), etc.

Finally, the host category contains all tables and views of the underlying DBMS (MS Access or SQL Server) metadata catalogs that *MatBase* accesses

### *MatBase* metadata catalog management

*MatBase* metadata catalog was designed and is managed only in (E)MDM. As for its development, a very small nucleus (SETS, FUNCTIONS, CONSTRAINTS, PROGRAMS, INF_RULES, OPERATORS, etc.) was implemented manually; then, all the rest was generated by this nucleus as its developers added metadata on the rest of the sets, functions, and constraints.

The main *MatBase* metadata management tasks are:

- Assist users in creating, updating, and deleting db schemes in both (E)MDM, E-RDM and RDM.
- Assist users in browsing, adding, updating and deleting data instances.
- Automatically generate standard MS Windows forms and their classes containing needed methods to enforce non-relational constraints.
- Automatically generate E-RDs from (E)MDM schemes and vice-versa.
- Import legacy dbs and generate corresponding (E)MDM schemes.
- Export (E)MDM schemes and their instances in both XML, HTML, DOCX, and PDF files.
- Assist users in detecting all existing keys for any object set [19].
- Detect and categorize all cycles in E-RDs [18].
- Assist users in analyzing E-RD cycles [12].
- Automatically detect incoherencies in constraint sets and assist users in removing them [4].
- Automatically detect and remove redundant constraint enforcement [4].
- Assist users in managing and running their Datalog programs [1,4].

### Conclusion

Data quality is a crucial dimension of today's data management. All constraints (which formalize business rules) that are governing the sub-universes modeled by dbs should be enforced in the corresponding dbs' schemas: otherwise, their instances might be implausible. As [15] puts it in its 10th rule (Data Integrity Is Its Own Reward) "each 1% data integrity failures will double the amount of time you spend troubleshooting them" and in its 11th one (The Data Integrity Tipping Point) "any database which contains 20% or more untrustworthy data is useless and will cost less to replace from source data than to fix". Not only in our opinion, today's businesses cannot be successful if their data is not almost 100% trustworthy.

Consequently, (E)MDM, a semantic approach to data modeling, provides very many constraint types, much more than any other data model, and *MatBase*, the prototype KDBMS built on it (as well as on the E-RDM and RDM), is offering them to db designers and users. This is done based on a complex metadata catalog made of over 370 tables and views. This catalog was designed, even partially implemented, and managed through the (E)MDM interface of *MatBase*.

### Assets from publication with us

- Prompt Acknowledgement after receiving the article
- Thorough Double blinded peer review
- Rapid Publication
- Issue of Publication Certificate
- High visibility of your Published work

**Website:** www.actascientific.com/
**Submit Article:** www.actascientific.com/submission.php
**Email us:** editor@actascientific.com
**Contact us:** +91 9182824667